\documentclass[letterpaper,sort&compress]{JHEP3}

\usepackage[dvips]{graphicx}
\usepackage{epsfig,multicol,bbm}
\usepackage{amsmath}
\usepackage{appendix}

\def\lsim{\mathrel{\hbox{\rlap{\hbox{\lower4pt\hbox{$\sim$}}}\hbox{$<$}}}}

\makeatletter

\newcommand{\Rmnum}[1]{\expandafter\@slowromancap\romannumeral #1@}
\makeatother

\title{Channeling in direct dark matter detection III:\\ channeling fraction in CsI crystals}

\author{Nassim Bozorgnia\\
Department of Physics and Astronomy, UCLA, 475 Portola Plaza, Los
  Angeles, CA 90095, USA\\
E-mail: \email{nassim@physics.ucla.edu}}

\author{Graciela B. Gelmini\\
Department of Physics and Astronomy, UCLA, 475 Portola Plaza, Los
  Angeles, CA 90095, USA\\
Email: \email{gelmini@physics.ucla.edu}}

\author{Paolo Gondolo\\
Department of Physics and Astronomy, University of Utah, 115 South 1400 East \#201,
  Salt Lake City, UT 84112, USA\\
  School of Physics, KIAS, Seoul 130-722, Korea\\
E-mail: \email{paolo@physics.utah.edu}}

\abstract{The channeling of the ion recoiling after a collision with a WIMP changes the ionization signal
in direct detection experiments, producing a larger signal scintillation or ionization than otherwise expected. We
give estimates of the fraction of channeled recoiling ions in CsI  crystals
using analytic models produced since  the 1960's and 70's  to describe channeling and blocking effects.}

\begin{document}

\section{Introduction}

Understanding the possibility of  having channeled ion recoils  in crystals is important in  direct dark matter experiments measuring ionization or scintillation signals~\cite{Drobyshevski:2007zj}-\cite{Avignone:2008cw}. These experiments  search for Weakly Interacting Massive Particles (WIMPs) composing the dark matter halo of our galaxy through the energy they deposit in collisions with nuclei within a crystal.
Channeling occurs when the nucleus that recoils after being hit by a dark matter particle moves off in a direction close to a symmetry axis or symmetry plane of the crystal. Channeled ions suffer a series of small-angle scatterings  with lattice nuclei that maintain them in the open ``channels"  between the rows or planes of lattice atoms. Thus  they penetrate much further into the crystal than in other directions and give 100\% of their energy to electrons (their quenching factor is $Q=1$), producing more scintillation and ionization than they would produce otherwise.  In scintillators like NaI (Tl) or CsI (Tl), channeling increases the observed  scintillation light output corresponding to a particular recoil energy. In dark matter searches, CsI (Tl) crystals are used by the KIMS collaboration~\cite{Kim-et-al-2003}.
 In this paper we give upper bounds to the geometric channeling fraction of recoiling ions in CsI crystals.

  We proceed  here in a similar manner as we did for NaI, a very similar crystal, in a previous paper~\cite{BGG-I} (we have also already considered channeling of recoiling ions in Si and Ge crystals in a different paper~\cite{BGG-II}).
 We use a continuum classical analytic model of channeling developed in the 1960's and 70's, in particular by Lindhard~\cite{Lindhard:1965}-\cite{Hobler}. In this model the discrete series of binary collisions of the propagating ion with atoms is replaced by a continuous interaction between the ion and uniformly charged strings or planes. The   screened atomic Thomas-Fermi potential is averaged over a direction parallel to a row or a plane. This averaged potential, $U$ or $U_p$, is considered to be uniformly smeared along the row or plane of atoms, respectively, which is a good approximation if the propagating ion interacts with many lattice atoms in the row or plane by a correlated series of many consecutive glancing collisions with lattice atoms. We are going to consider just one row or one plane, which simplifies the calculations and is
correct except at the lowest energies we consider.

In order for the scattering  to happen at small enough angles so that channeling is maintained, the propagating ion must not approach a string or plane closer than a critical distance $r_c$  or $x_c$ respectively ($r$ is the  transverse distance to the string and $x$ is the distance perpendicular to the plane). The critical distance depends on the energy $E$ of the ion and on the temperature of the crystal. Ions which start their motion close to the center of a channel (for example ions incident upon the crystal from outside),  far from a row or plane, are channeled if the angle their trajectory makes with the row or plane is smaller than a critical angle, $\psi_c$ that depends on the critical distance of approach $r_c$  or $x_c$, and are not  channeled otherwise. Nuclei ejected from their lattice site by WIMP collisions are initially part of a row or plane, so they start their motion  from lattice sites or very close to them. This means that ``blocking effects'', namely large-angle interactions with the nuclei in the lattice sites directly in front of the recoiling nucleus site,  are important. In fact, as argued originally by Lindhard~\cite{Lindhard:1965}, in a perfect lattice and in the absence of energy-loss processes, the probability that a particle starting from a lattice site is channeled would be zero.  This is what Lindhard called the ``Rule of Reversibility." However, any departure of  the actual lattice from a perfect lattice due to vibrations of the atoms,  which are always present, violate the conditions of this argument and allow for some of the recoiling lattice nuclei to be channeled.

There are several good analytic approximations of the screened Thomas-Fermi potential and each leads to a different expression for the transverse continuum string and plane potentials, $U(r)$ and $U_p(x)$ respectively. As in Ref~\cite{BGG-I} here we use Lindhard's expression, because it is the simplest and allows to find analytical expressions for the quantities we need. The
  transverse averaged continuum potential of a string as a function of  $r$, relevant for  axial channeling,  was approximated by Lindhard~\cite{Lindhard:1965} as
\begin{equation}
U(r)=E \psi_{1}^2 \, \frac{1}{2}\ln\left(\frac{C^2 a^2}{r^2}+1\right),
\label{axial-pot}
\end{equation}
where  $C$ is a constant found experimentally to be $C\simeq\sqrt{3}$~\cite{Lindhard:1965} and $\psi_{1}^2=2Z_{1}Z_{2}e^2/(E d)$. Here $Z_1$ and $Z_2$ are the atomic numbers of the recoiling and lattice nuclei respectively, $d$ is the spacing between atoms in the row, $a$ is the Thomas-Fermi screening distance, $a= 0.4685 {\text {\AA} } (Z_1^{1/2} + Z_2^{1/2})^{-2/3} $~\cite{Barrett:1971, Gemmell:1974ub} and   $E= Mv^2/2$ is the kinetic energy of the propagating ion.
  In our case, $E$ is the recoil energy imparted to the ion after a collision with a WIMP,
\begin{equation}
E = \frac{|\vec{\bf q}|^2}{2M} ,
\end{equation}
where $\vec{\bf q}$ is the recoil momentum. The string of crystal atoms is at $r=0$.
The transverse averaged continuum potential of a plane of atoms, relevant for planar channeling,  given by Lindhard~\cite{Lindhard:1965} as a function of  $x$ is
\label{eq:planar}
\begin{equation}
U_p(x)=E \psi_a^2\left[\left(\frac{x^2}{a^2}+C^2\right)^\frac{1}{2}-\frac{x}{a}\right],
\label{planar-pot}
\end{equation}
where $\psi_a^2=2\pi n Z_1 Z_2 e^2 a/E$, $n= N d_{pch}$ is the average number of atoms per unit area, $N$ is the atomic density and $d_{pch}$ is the width of the planar channel. Also, the axial channel width $d_{\rm ach}$ is defined in terms of the interatomic distance $d$ as $d_{\rm ach}= 1/ \sqrt{N d}$, with $N$ the atomic density. The plane is at $x=0$.
Examples of axial and planar continuum potentials for Cs ions propagating in the $<$100$>$ axial and \{100\} planar channels of a CsI crystal are shown in Fig.~\ref{U}. The potentials for Cs and I ions are practically identical, because $Z_{Cs} \simeq Z_{I}$ (see Appendix A).
\FIGURE{\epsfig{file=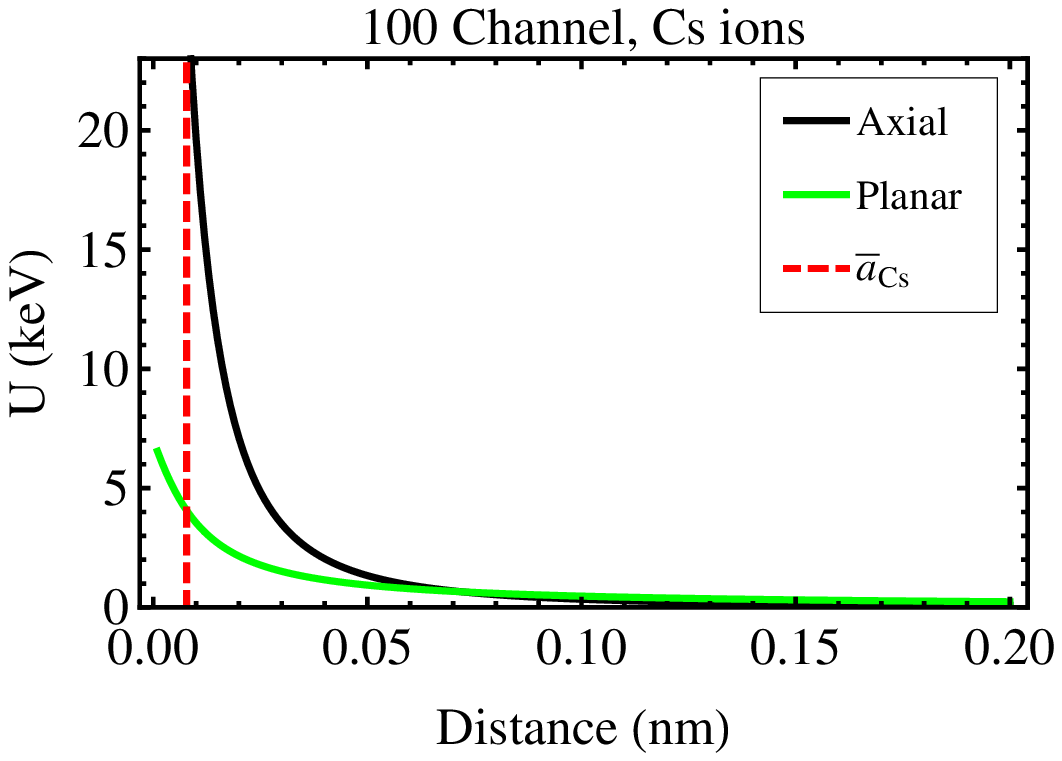,height=210pt}
        \caption{Continuum axial (black) and planar (green/gray) potentials for Cs ions, propagating  in the $<$100$>$ axial and \{100\} planar channels of a CsI crystal. The screening radius shown as a vertical line is $\bar{a}_{\rm Cs}=0.007785$ nm (see App. A).}
	\label{U}}

  For a ``static lattice," that here means a perfect lattice in which all vibrations are neglected, the critical distances of approach $r_c$ and $x_c$ are given in the  Eqs.~\ref{rcrit} and \ref{ourxcrit}, expressions that were derived in Ref~\cite{BGG-I} . The critical distance for axial channeling is
\begin{equation}
r_c(E) =Ca\, \sqrt{\frac{2}{3} \left[ \sqrt{1+z} \, \cos\!\left( \frac{1}{3} \arccos\frac{(1-3z/2)}{(1+z)^{3/2}} \right) - 1\right] },
\label{rcrit}
\end{equation}
where $z=9 E_1 d^2/(8 E C^2  a^2)$. For planar channeling we will follow the procedure of defining a  ``fictitious string" introduced by Morgan and Van Vliet~\cite{Morgan-VanVliet, Hobler}.
They reduced the problem of scattering from a plane of atoms to the scattering of an equivalent row of atoms contained in a strip of a certain width  centered on the projection of the ion path onto the plane of atoms. Thus,
\begin{equation}
x_c(E) \equiv \bar{r}_c(E),
\label{ourxcrit}
\end{equation}
where $\bar{r}_c(E)$ is the critical distance obtained from  Eq.~\ref{rcrit} for the fictitious string. Along the fictitious row, the characteristic distance $\bar{d}$ between atoms needs to be estimated using data or simulations which are not available for a CsI crystal. As explained in Ref.~\cite{BGG-I}, the choice of $\bar{d}$  equal to the average  interdistance of atoms in the plane $d_p$, i.e. $\bar{d}=d_p$, yields a lower bound on $x_c$, which translates into an upper bound on the fraction of channeled recoils into planar channels.

So far we have been considering static strings and planes, but the atoms in a crystal are actually vibrating with a characteristic (one dimensional rms) amplitude of vibration $u_1(T)$ which increases with the temperature $T$. In principle there are modifications to the continuum potentials due to thermal effects, but we  take into account thermal effects  in the crystal through a modification of the critical distances found originally by Morgan and Van Vliet~\cite{Morgan-VanVliet}  and later by Hobler~\cite{Hobler} to provide good agreement with simulations and data. For axial channels it consists of taking the temperature corrected critical distance $r_c(T)$  to be,
\begin{equation}
r_c(T)= \sqrt{r^2_c(E) + [c_1 u_1(T)]^2},
\label{rcofT}
\end{equation}
where the dimensionless factor $c_1$ in different references is a number between 1 and 2 (see  e.g. Eq. 2.32 of Ref.~\cite{VanVliet} and
Eq. 4.13 of the 1971 Ref.~\cite{Morgan-VanVliet}). For planar channels, following Hobler ~\cite{Hobler} we use a  similar equation
\begin{equation}
x_c(T)= \sqrt{x^2_c(E) + [c_2 u_1(T)]^2},
\label{xcofT}
\end{equation}
where again $c_2$ is a number between 1 and 2 (for example Barret~\cite{Barrett:1971} finds $c_2 = 1.6$ at high energies, and Hobler~\cite{Hobler}  uses $c_2 = 2$). We will mostly use $c_1=c_2=1$ in the following, to try to produce upper bounds on the channeling fractions.

 In Appendix B it is shown that the variation of the lattice size with temperature, characterized by the variation of the lattice constant $a_{\rm lat}$ with temperature, has a negligible effect on the channeling fractions. This is why we ignore this effect (not only in this paper but also in our previous papers~\cite{BGG-I, BGG-II}).

We use the Debye model to account  for the vibrations of the atoms in a crystal.  The one dimensional rms vibration amplitude $u_1$ of the atoms in a crystal in this model  is~\cite{Gemmell:1974ub, Appleton-Foti:1977}
\begin{equation}
u_1(T)=12.1 \, \text{\AA} \, \left[\left(\frac{\Phi(\Theta/T)}{{\Theta/T}} + \frac{1}{4}\right)(M\Theta)^{-1}\right]^{1/2},
\label{vibu1}
\end{equation}
where $M$ for a compound is the average atomic  mass (in amu), i.e.  for CsI,  $M= (M_{\rm Cs} + M_{\rm I})/2$, $\Theta$ and $T$ are the Debye temperature and the temperature of the crystal (in K), respectively, and $\Phi(x)=\frac{1}{x}\int_{0}^{x}{t dt/(e^t -1)}$ is the Debye function. With $M_{\rm Cs}=132.9$ amu and $M_{\rm I}=126.9$ amu, then $M=129.9$ amu.  We take the Debye temperature of CsI to be $\Theta=125\;$K~\cite{Gemmell:1974ub, Sharko-Botaki-1971}. The  one dimensional rms  vibration amplitude $u_1$  in CsI is plotted in Fig.~\ref{figu1} as a function of the temperature $T$.  The crystals in the KIMS experiment were kept at 0 $^\circ$C in 2007~\cite{Lee-2007}. Currently the operating temperature of the crystals is 20 $^\circ$C~\cite{Kims-private}. The vibration amplitude is $u_1=0.0141$ nm at 0 $^\circ$C, and $u_1=0.0146$ nm  at 20 $^\circ$C.
\FIGURE{\epsfig{file=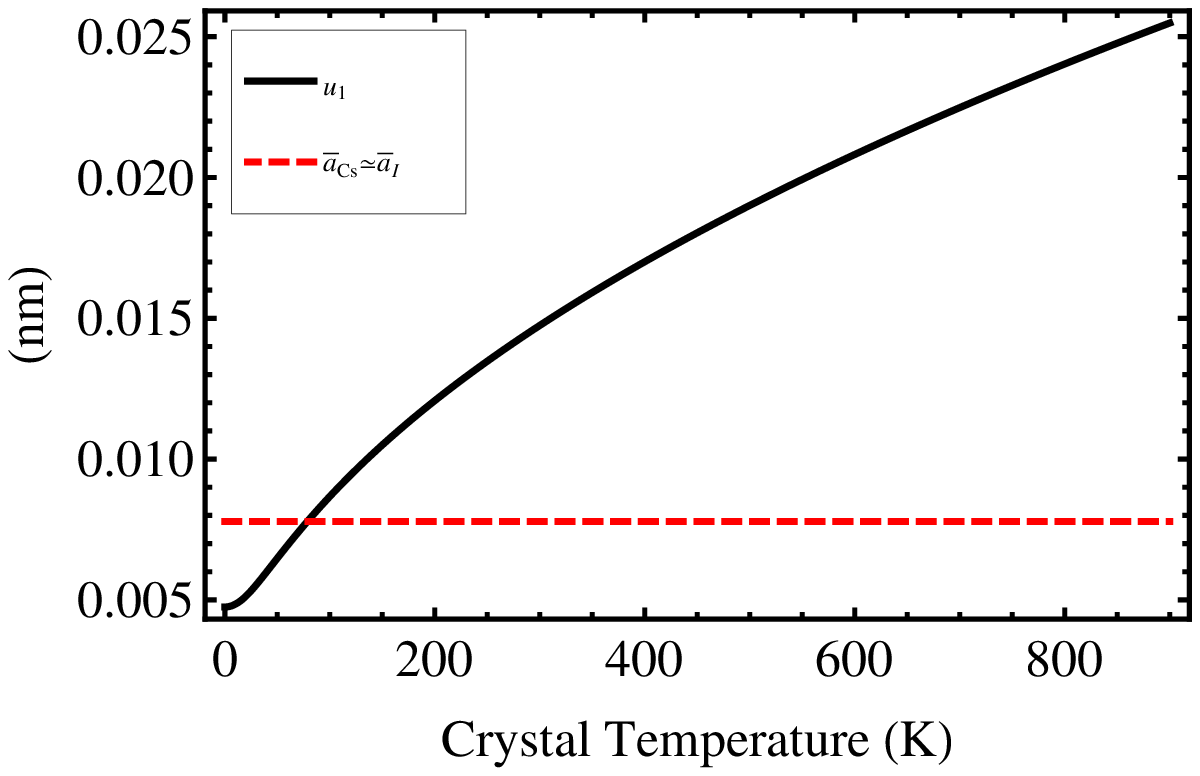,height=210pt}
        \caption{Plot of $u_1(T)$ for CsI (Eq.~\ref{vibu1} with $M= (M_{\rm Cs} + M_I)/2$).}
	\label{figu1}}

 Using the temperature corrected critical distances of approach $r_c(T)$ and $x_c(T)$ (Eqs.~\ref{rcofT} and \ref{xcofT}) or  the static lattice critical distances $r_c$ and $x_c$ (Eqs.~\ref{rcrit} and \ref{ourxcrit}), we obtain the corresponding critical axial and planar channeling angles $\psi_c$ (see Ref.~\cite{BGG-I} for details). Examples are shown in  Figs.~\ref{rc-CsI100-DiffT-c1} to \ref{rc-CsI100-DiffT-c2}, for $c_1=c_2=c$ and $c=1$ or  $c=2$ as indicated.
\FIGURE[h]{\epsfig{file=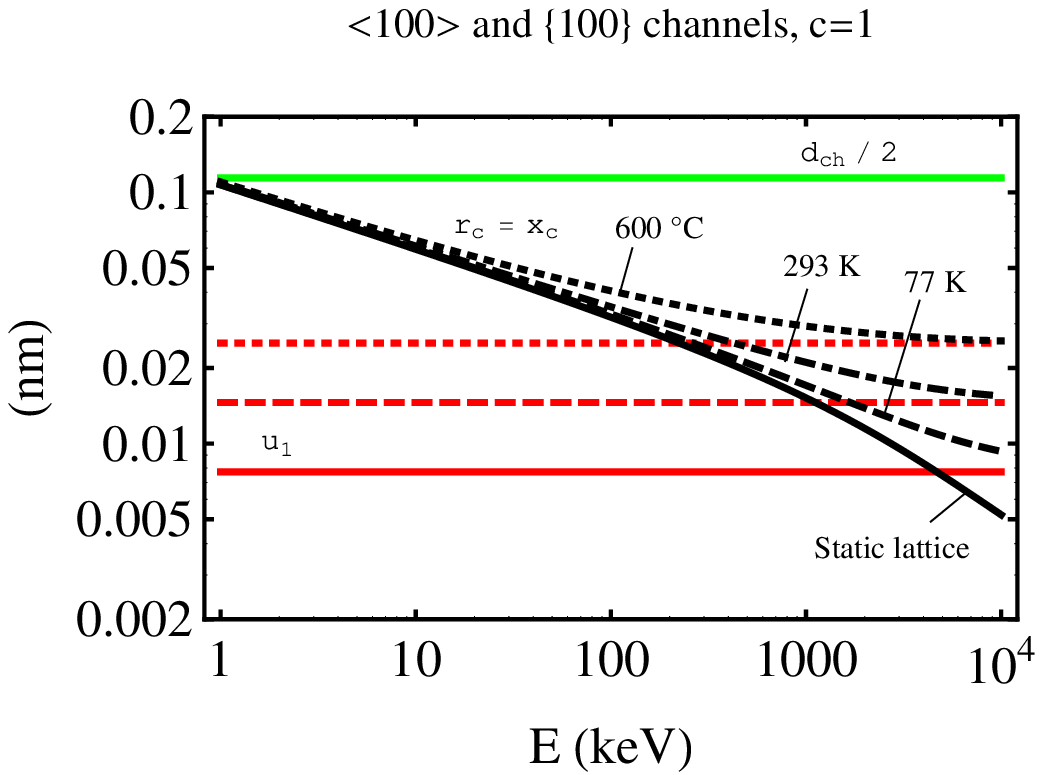,height=160pt}
\epsfig{file=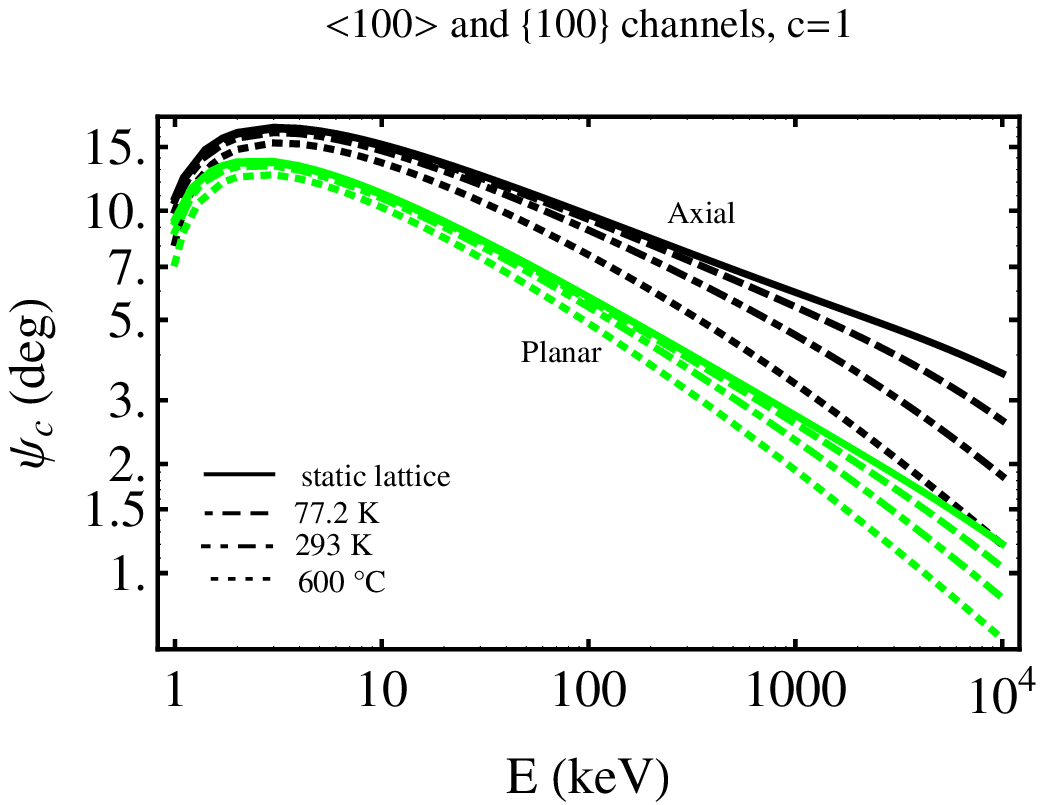,height=160pt}\\
        \vspace{-0.5cm}\caption{Static (green) and temperature corrected with $c_1=c_2=c=1$ (black) (a) critical distances of approach (and $u_1(T)$ in red) and (b) the corresponding critical channeling angles, as a function of the energy of propagating Cs or I ions (they are practically the same for both) in the $<100>$ axial (black) and \{100\} planar (green) channels. Here $d_{\rm ach}=d_{\rm pch}$.}%
	\label{rc-CsI100-DiffT-c1}}
\FIGURE{\epsfig{file=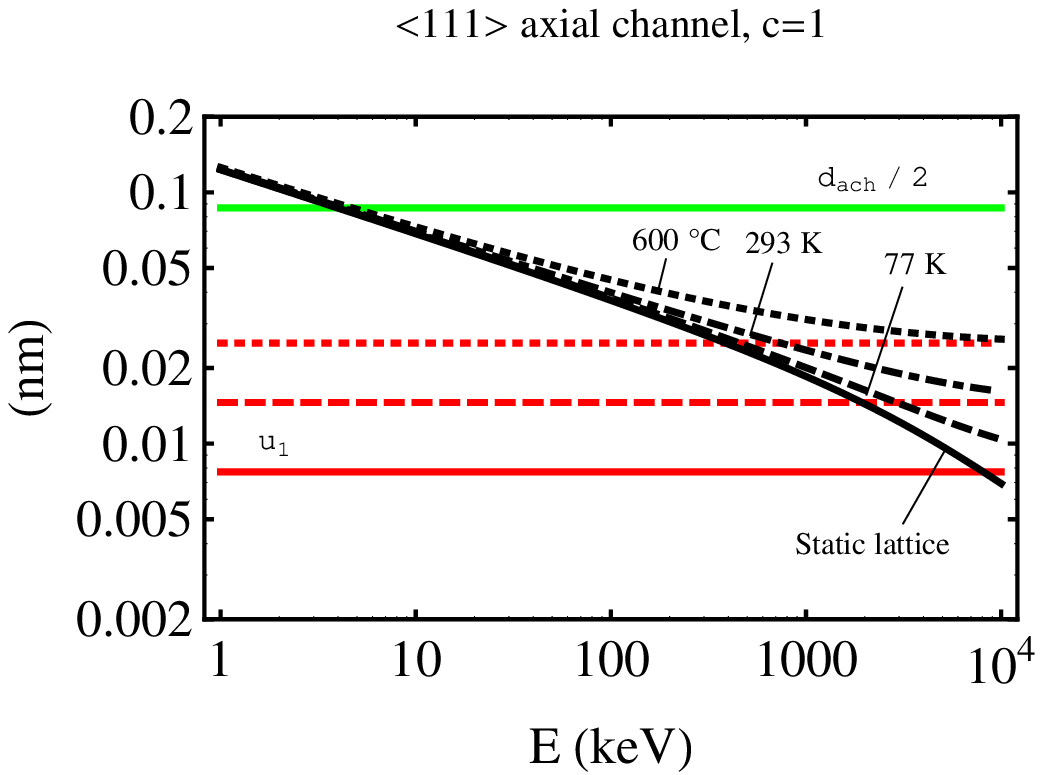,height=160pt}
\epsfig{file=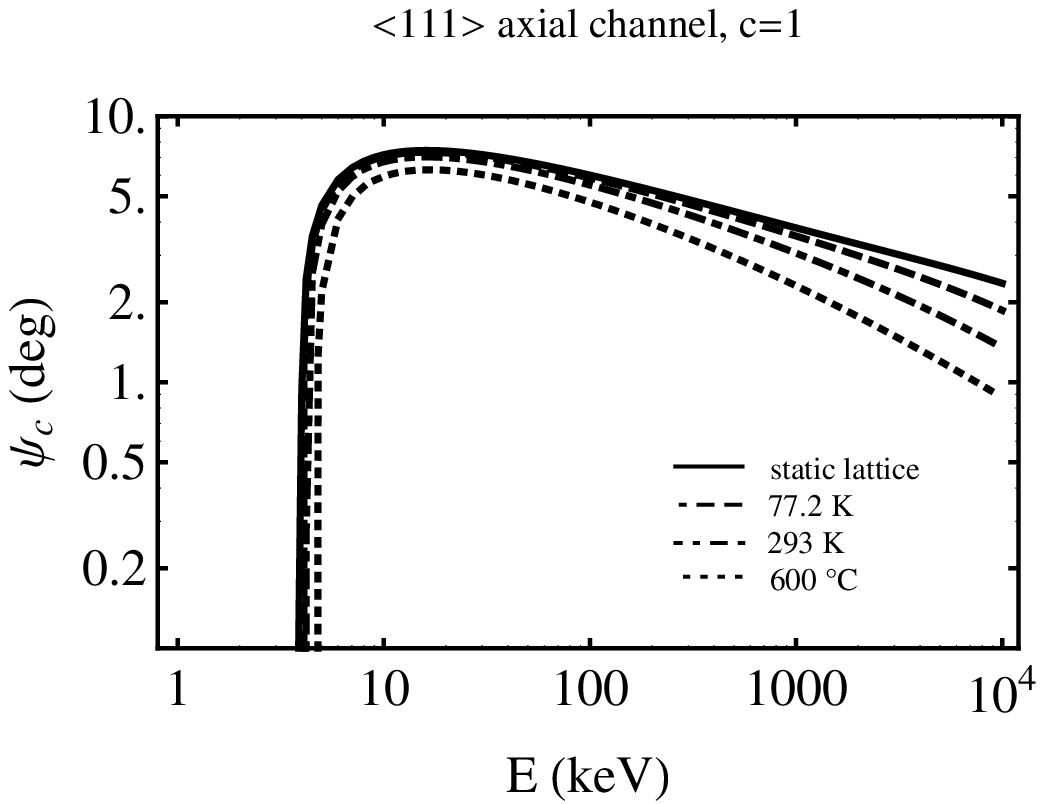,height=160pt}\\
        \vspace{-0.5cm}\caption{Same as Fig.~\ref{rc-CsI100-DiffT-c1} but for the $<$111$>$ axial channel.}%
	\label{rc-CsI111-DiffT-c1}}

Fig.~\ref{rc-CsI100-DiffT-c1} clearly shows how the critical distances and angles change with temperature for a Cs or I ion propagating in the  $<$100$>$ axial and \{100\} planar channels of a CsI crystal, with temperature effects computed with $c_1=c_2=c=1$. At small energies the static critical distance of approach is much larger than the vibration amplitude, so temperature corrections are not important. As the energy increases, the static critical distance of approach decreases, and when it becomes negligible with respect to the vibration amplitude $u_1$, the temperature corrected critical distance $r_c$ becomes equal  to $c_1 u_1$. In this case, since $u_1(T)$ increases with $T$, the critical distance $r_c \simeq c_1 u_1$ becomes larger with $T$, and therefore the critical channeling angle becomes smaller. Notice that for the 100 channels, the widths of axial and planar channels are the same, $d_{\rm ach}=d_{\rm pch}$ and $r_c=x_c$.

Fig.~\ref{rc-CsI111-DiffT-c1} shows the same effects for the  $<$111$>$ axial channel. In this channel,  the critical distance (the minimum distance to a string to maintain channeling) becomes larger than the radius of the channel at energies  below a few keV, shown in the figures. This means that nowhere in the channel an ion can be far enough from the string of lattice atoms for channeling to take place  (thus the critical channeling angle is zero). The exact calculation of the energy at which this happens would require considering the effect of more than a single row of atoms (which we do not do here) thus our results at these low energies are only approximate. Notice  that for the 111 channels,  the $<$111$>$ axial and   \{111\} planar channels do not have the same widths, $d_{\rm ach}\not= d_{\rm pch}$, and we only show  the critical distance and angles for the axial channel.

Figs.~\ref{rc-CsI100-DiffT-c2}(a) and \ref{rc-CsI100-DiffT-c2}(b) show the static and $T$-corrected critical distances and angles repectively at several temperatures for traveling Cs or I ions in the 100 axial and planar channels with $c_1=c_2=c=2$.
\FIGURE{\epsfig{file=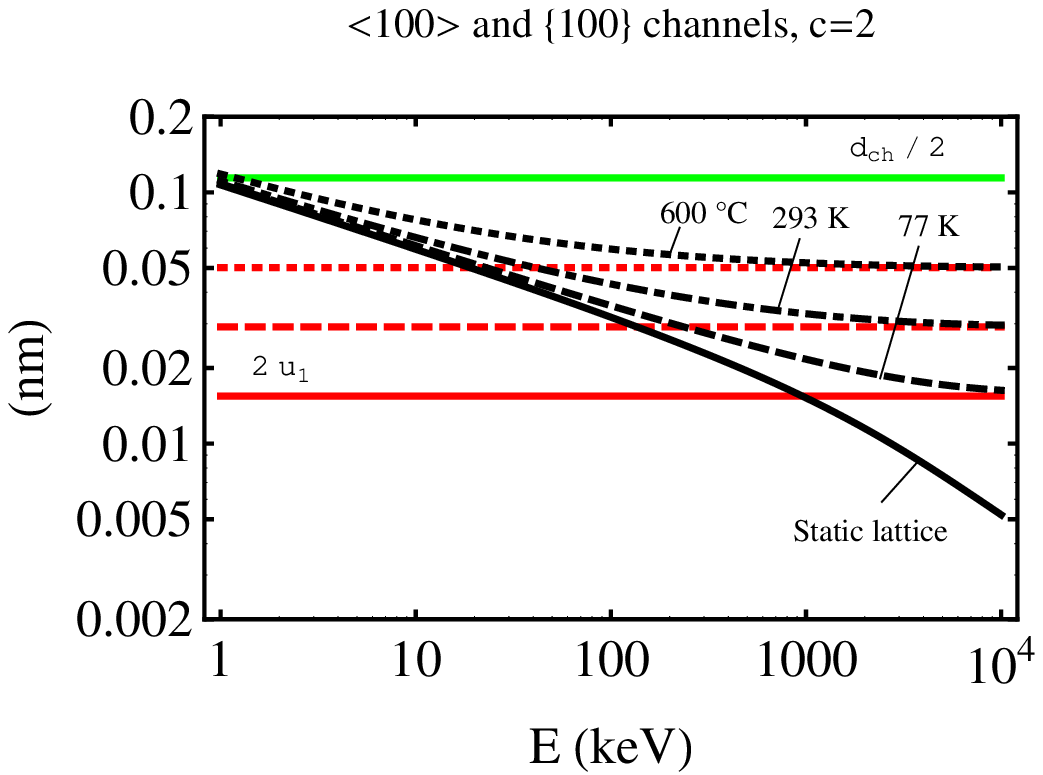,height=160pt}
\epsfig{file=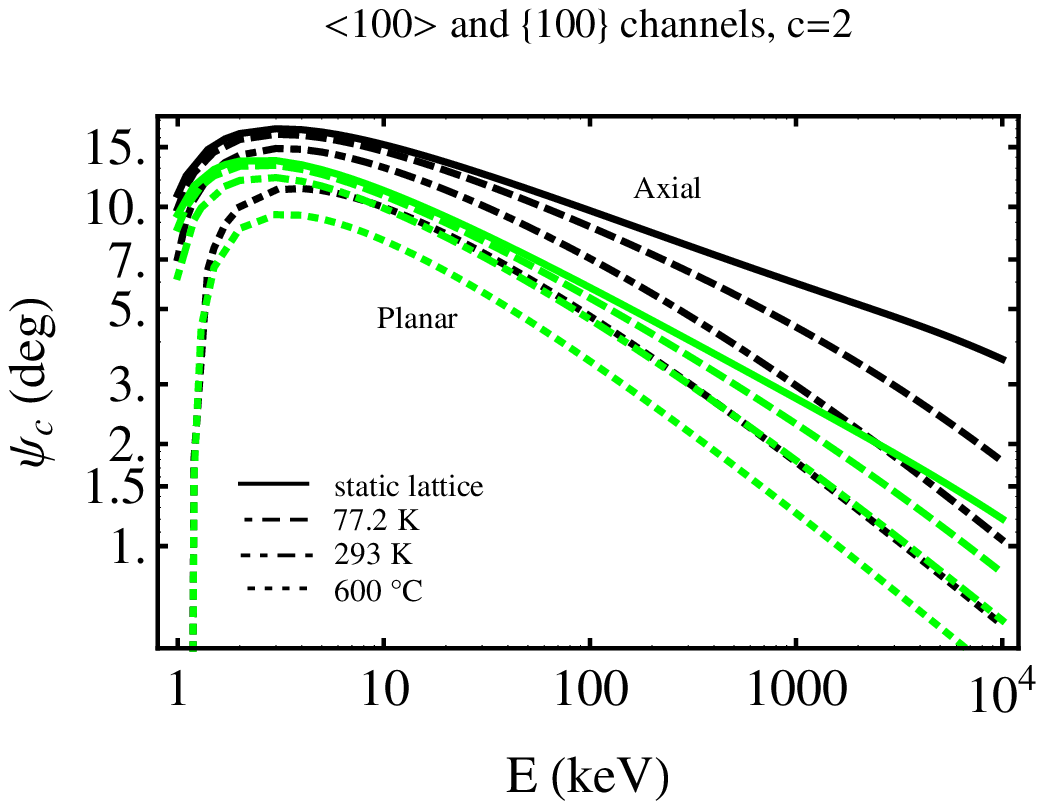,height=160pt}\\
        \vspace{-0.5cm}\caption{Same as Fig.~\ref{rc-CsI100-DiffT-c1} but with $c_1=c_2=c=2$.}%
	\label{rc-CsI100-DiffT-c2}}

\section{Channeling fractions}

In  our model, a recoiling ion is channeled if the collision with a WIMP happened at a distance large enough from  the string  or plane from which the ion was expelled. Namely, channeling happens if the initial position of the recoiling motion is $ r_i > r_{i,\rm min}$ or $ x_i> x_{i,\rm min}$ for an axial or planar channel respectively. Here
$r_{i,\rm min}$ and $x_{i,\rm min}$ depend  on the critical distance,  $r_c$ or $x_c$
 (and through it  on the energy $E$ and the temperature $T$) and on the angle $\phi$ the initial ion's momentum makes with the string or plane. In Ref.~\cite{BGG-I} we obtained the following expressions for both distances:
\begin{equation}
r_{i,\rm min} (E, \phi) = \frac{C a}{\sqrt{\left( 1+\frac{C^2a^2}{r_{c}^2} \right) \, \exp\!\left(-{2 \sin^2\phi}/{\psi_1^2} \right) -1 }} - d \tan\phi,
\end{equation}
with $r_c$  given in Eq.~\ref{rcofT} and
\begin{equation}
x_{i, \rm min}(E, \phi)=\frac{a}{2}\,\frac{C^2-\left[\sqrt{\frac{x_c^2}{a^2}+C^2}-\frac{x_c}{a}-{\sin^2\phi}/{\psi_a^2}\right]^2}{\left[\sqrt{\frac{x_c^2}{a^2}+C^2}-\frac{x_c}{a}-{\sin^2\phi}/{\psi_a^2}\right]} - d_p \tan\phi,
\end{equation}
where $x_c$ is given in Eq.~\ref{xcofT}.

We take the initial distance distribution of the colliding atom to be a Gaussian with  a  one dimensional dispersion $u_1$, and to obtain the probability of channeling for each individual channel  we integrate the Gaussian between the minimum initial distance and infinity (a good approximation to the radius of the channel; see Ref~\cite{BGG-I} for details). The dependence of these probabilities on the critical distances enter in the argument of an exponential or an erfc function. Thus any uncertainty in our modeling of the critical distances becomes exponentially enhanced in the channeling fraction.  This is the major difficulty of the analytical approach we are following.

In order to obtain the total geometric channeling fraction we need to sum over all the individual channels we consider.
Taking only the channels with lowest crystallographic indices, 100, 110 and 111, we have a total of 26 axial and planar channels, as explained in Appendix A of Ref.~\cite{BGG-I} (CsI and NaI have the same crystal structure).  Here ``geometric channeling fraction'' refers to assuming that the distribution of recoil directions is isotropic. In reality, in a dark matter direct detection experiment, the distribution of recoil directions is expected to be peaked in the direction of the average WIMP flow. The integral over initial directions is
computed using HEALPix~\cite{HEALPix:2005} (see Appendix B of Ref.~\cite{BGG-I}).

Fig.~\ref{Our-HEALPIX} shows upper bounds to the channeling probability  computed for each initial recoil direction direction $\hat{\bf q}$ and plotted on a sphere using the HEALPix pixelization for  (a)  a $E= 200$ keV  and  (b) a 1 MeV Cs ion at 20 $^\circ$C with  $c_1=c_2=1$ assumed for the temperature effects. The red, pink, dark blue and light blue colors indicate a channeling probability of 1, 0.625, 0.25 and zero, respectively. The results are practically identical for an I ion.

Fig.~\ref{OneChannel} shows upper bounds to the channeling fractions of Cs recoils for individual channels,  for T $=20 ^\circ$C  and assuming $c_1=c_2=1$. The black and green (or gray) lines correspond to single axial and planar channels respectively. The upper bounds of the channeling fractions of planar channels are more generous than those of axial channels because of our choice of $x_c$ in Eq.~\ref{ourxcrit}. This does not mean that planar channels are dominant in the actual channeling fractions.
\FIGURE{\epsfig{file=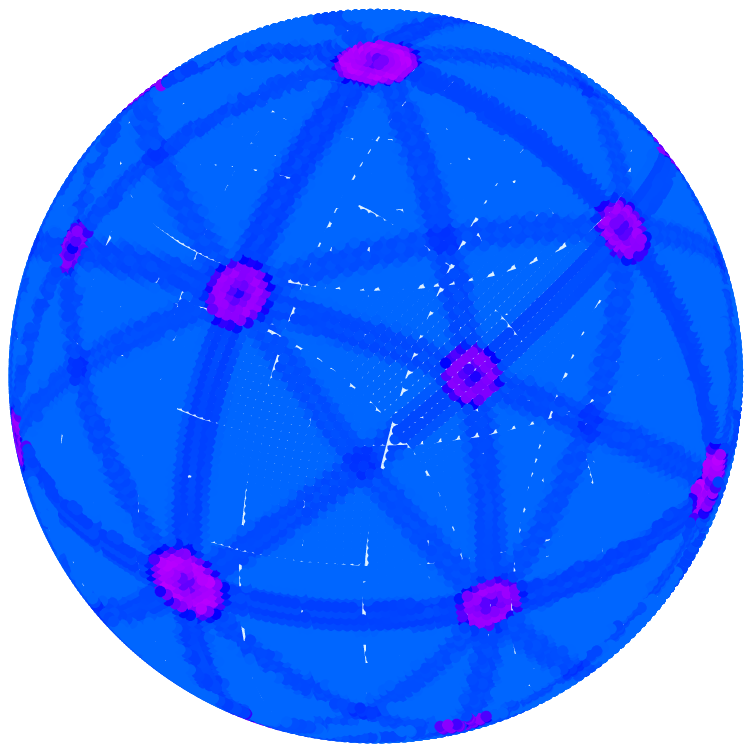,height=170pt}
\epsfig{file=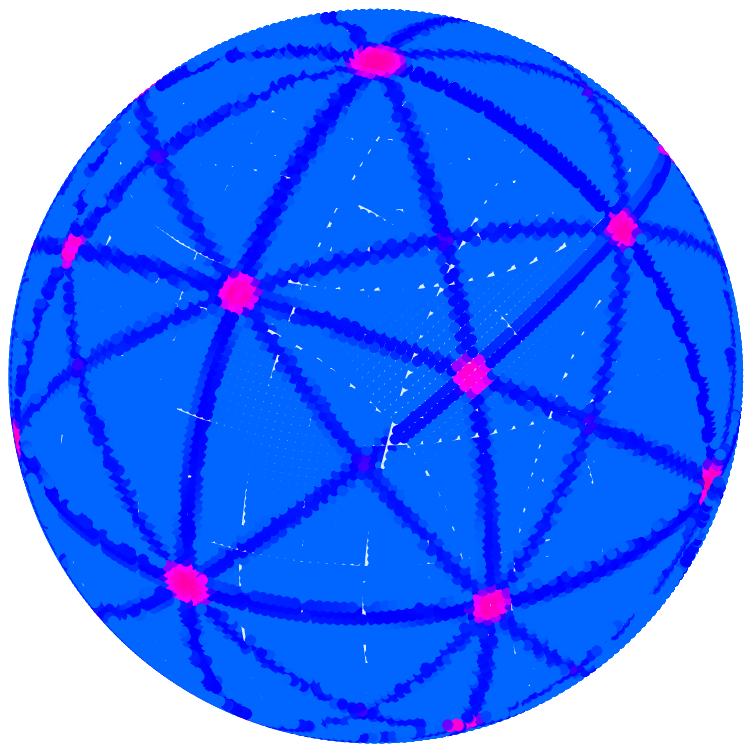,height=170pt}
        \vspace{-0.5cm}\caption{Upper bounds on the channeling probability of a Cs ion (for an I ion the figure would be practically identical ) as function of the initial recoil direction for a (a) 200 keV and (b) 1 MeV  recoil energy  at 20 $^\circ$C (with $c_1=c_2=1$). The probability is computed for each direction and plotted on a sphere using the HEALPix pixelization. The red, pink, dark blue and light blue colors indicate a channeling probability of 1, 0.625, 0.25 and zero, respectively.}%
	\label{Our-HEALPIX}}
\FIGURE{\epsfig{file=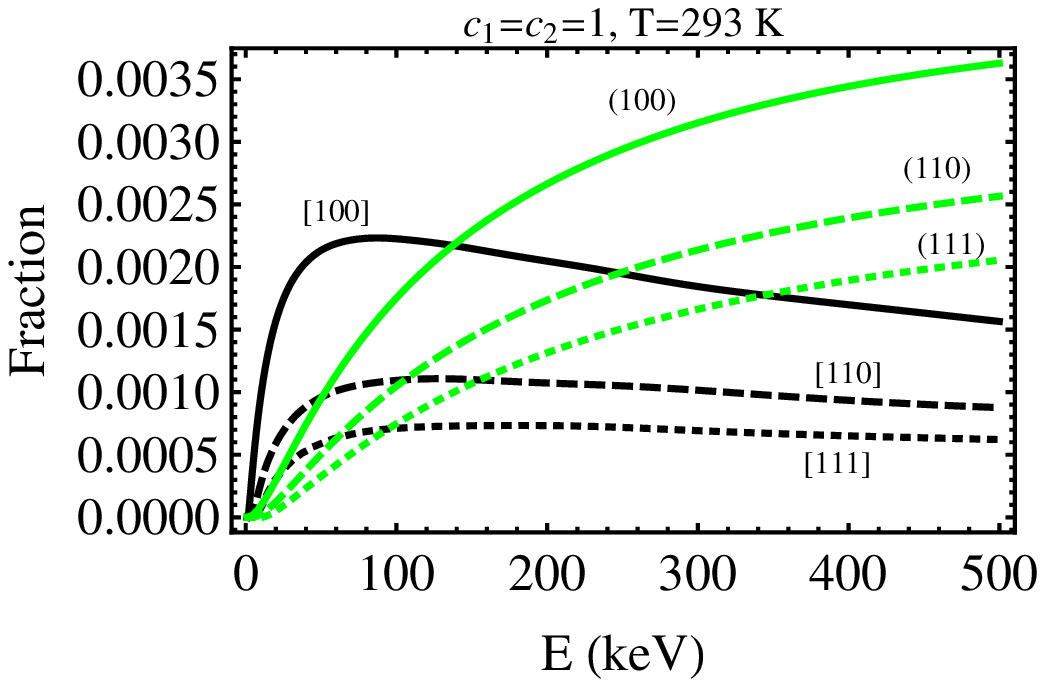,height=210pt}
        \caption{Upper bounds on the channeling fractions of Cs recoils as a function of the recoil energy $E$ when only one channel is open, for $T=293$ K with temperature corrections included in the critical distances with the coefficients $c_1=c_2=1$. Black and green/gray lines correspond to axial and planar channels respectively. Solid, dashed, and dotted lines are for 100, 110, and 111 channels respectively. The corresponding figure for an I ion would be practically identical.}%
	\label{OneChannel}}
\FIGURE{\epsfig{file=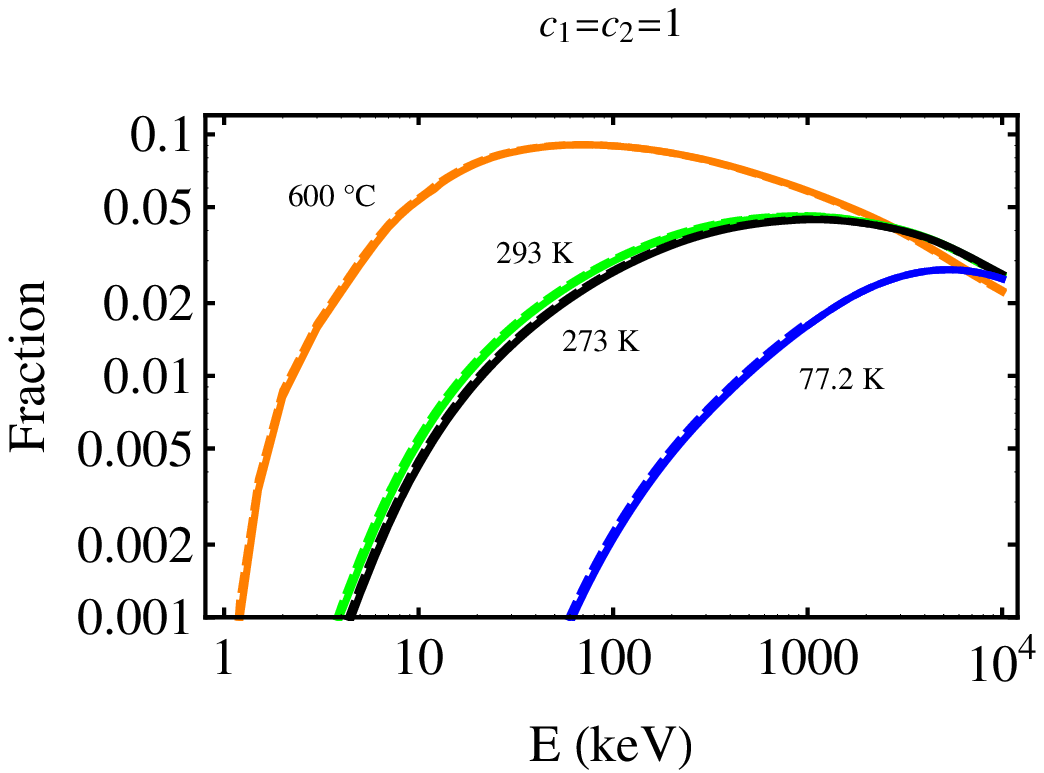,height=210pt}
        \caption{Upper bounds on the channeling fraction of Cs (solid lines) and I (dashed lines) recoils as a function of the recoil energy $E$ for $T=600$ $^\circ$C (orange/medium gray), 293 K (green/light gray), 273 K (black), and 77.2 K (blue/dark gray)  in the approximation of $c_1=c_2=1$ without dechanneling.}%
	\label{FracCsI-DiffT-c1}}

\FIGURE{\epsfig{file=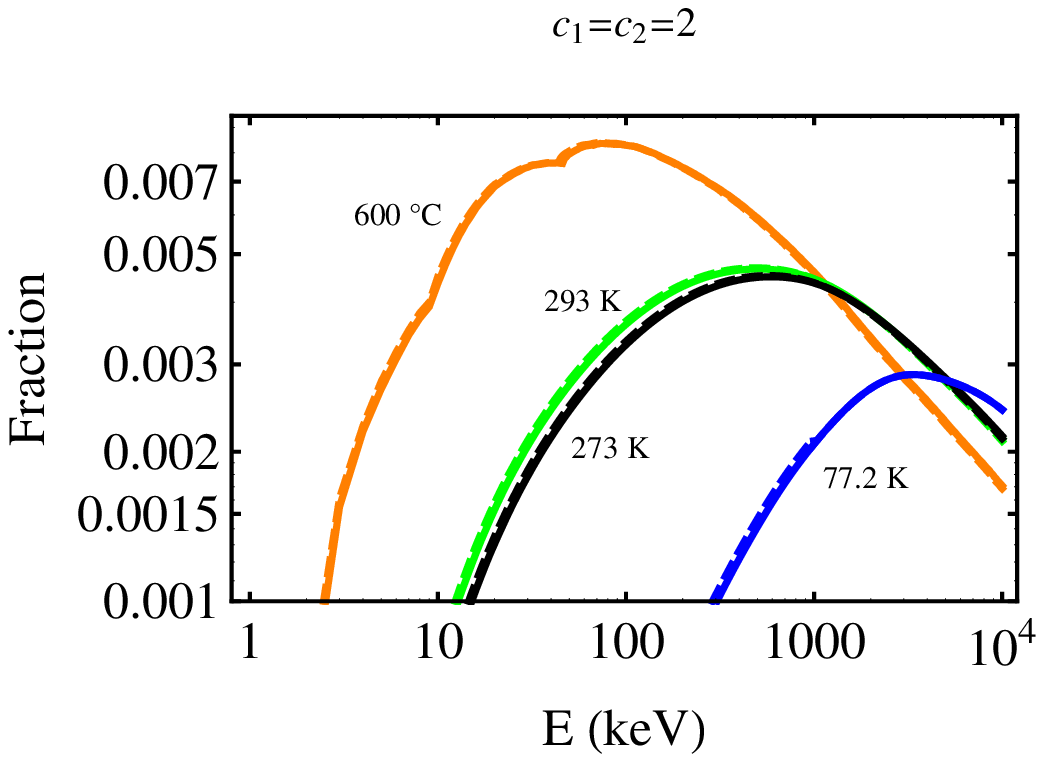,height=210pt}
        \caption{Same as Fig.~\ref{FracCsI-DiffT-rigid} but for $c_1=c_2=2$.}%
	\label{FracCsI-DiffT-c2}}

\FIGURE{\epsfig{file=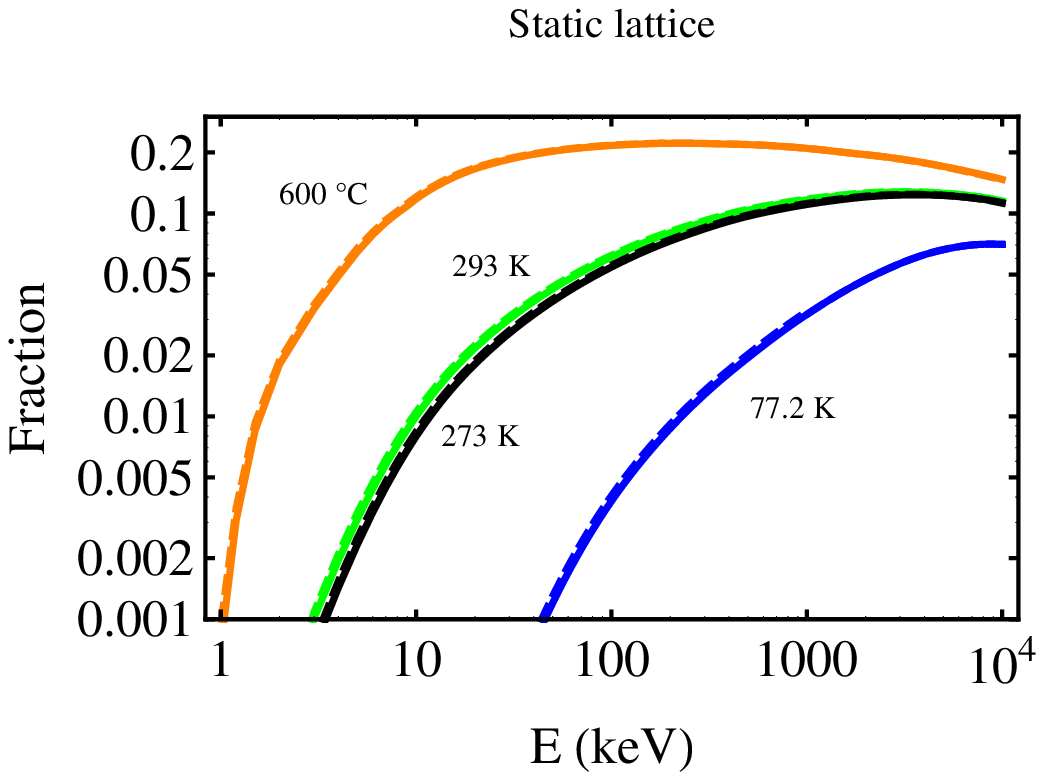,height=210pt}
        \caption{Same as Fig.~\ref{FracCsI-DiffT-rigid} but for $c_1=c_2=0$ (static lattice),  which provides an extreme upper bound (any larger values of $c_1$ and $c_2$, which can reasonably be as large as 2, yield smaller factions).}%
	\label{FracCsI-DiffT-rigid}}
Upper bounds to the geometric channeling fractions of Cs and I ions as  function of the  recoil energy are shown in Figs.~\ref{FracCsI-DiffT-c1} and \ref{FracCsI-DiffT-c2} with thermal effects taken into account with  $c_1=c_2=1$ and $c_1=c_2=2$, respectively.

 Notice that we have not included here any dechanneling due to the presence of impurities in the crystal (such as Tl atoms), which would decrease the channeling fractions presented. We intend to return to this issue in a later paper. As we see in Fig.~\ref{FracCsI-DiffT-c1} and \ref{FracCsI-DiffT-c2}, when no dechanneling is taken into account, the channeling fraction
increases with energy, reaches a maximum at a certain energy and then decreases as the energy increases. This  maximum  occurs because the critical distances decrease with the ion energy $E$, which makes channeling more
probable, while the critical angles also decrease with $E$, which makes channeling less probable. At low $E$ the critical distance effect dominates, and at large $E$ the critical angle effect dominates.

 As shown in Fig.~\ref{FracCsI-DiffT-c1}, the channeling fraction for CsI (Tl) is never larger than 5\% at 293 K (with $c1 = c2 = 1$) and the maximum fraction happens at around 1 MeV. This is comparable to the channeling fraction of Na (Tl) which is also never larger than 5\%, but in the case of Na (Tl) the maximum happens at 100's of keV (see Fig.~14(a) of Ref.~\cite{BGG-I}). For Si and Ge (with $c1 = c2 = 1$) the channeling fractions reach about 1\% and the maximum happens at 100's of keV for Si and at around 1 MeV for Ge (see Fig.~18 of Ref.~\cite{BGG-II}).
 Fig.~\ref{FracCsI-DiffT-c2}, shows  that the maximum channeling fraction for Cs or I recoils at 293 K  would be below 0.5\%  if $c_1=c_2=2$ instead. However, since we do not know which are the correct values of the crucial parameters $c_1$ and $c_2$ for CsI, we could ask ourselves how the upper bounds on channeling fractions would change if the values of these parameters would be smaller than 1 (please recall that for other materials and propagating ions the values of these parameters were found to be between 1 and 2). The values of $c_1$ and $c_2$ cannot be smaller than zero, thus  Fig.~\ref{FracCsI-DiffT-rigid} shows our most generous  upper bounds on the geometric channeling fraction, obtained by setting $c_1=c_2=0$, namely by neglecting thermal vibrations of the lattice (which make the channeling fractions smaller as $T$ increases) but including the thermal vibrations of the nucleus that is going to recoil (which make the channeling fraction larger as $T$ increases). Although it is physically inconsistent to take only the temperature effects on the initial position but not on the lattice,   we do it here because using $c_1=c_2=0$, namely a static lattice, provides an upper bound on the channeling probability with respect to that obtained using any other non-zero value of $c_1$ or $c_2$. Even in this case, the channeling fractions at 293 K cannot be larger than 10\%.

 To conclude, let us remark that the analytical  approach used here can successfully describe qualitative features of the channeling and blocking effects, but should be complemented by data fitting of parameters and by simulations to obtain a good quantitative description too.  Thus our results should in the last instance be checked by using some of the many sophisticated Monte Carlo simulation programs implementing the binary collision approach or mixed approaches.

\begin{acknowledgments}
N.B. and G.G. were supported in part by the US Department of Energy Grant
DE-FG03-91ER40662, Task C.  P.G. was  supported  in part by  the NFS
grant PHY-0756962 at the University of Utah. We would like to thank Prof.~Sun-Kee Kim for providing us with important information about the KIMS experiment.

\end{acknowledgments}

\appendix
\addappheadtotoc

\section{Crystal structure and other data for CsI}

CsI is a diatomic compound that has two interpenetrating face-centered cubic (f.c.c.) lattice structures displaced by half of a lattice constant with 8 atoms per unit cell. The lattice constant of CsI crystal is $a_{\rm lat}=0.45667$ nm at room temperature (Table 3.4 in Appleton and Foti~\cite{Appleton-Foti:1977}). The temperature dependence of $a_{\rm lat}$ is explained in Appendix B.

The atomic mass and atomic number of Cs and I are $M_{\rm Cs}=132.9$ amu, $M_{\rm I}=126.9$ amu, $Z_{\rm Cs}=55$ and $Z_I=53$.

With respect to  the Thomas-Fermi screening distance,  for Cs recoils from a mixed row or plane we use the average
\begin{equation}
 \bar{a}_{\rm Cs}=(a_{\rm CsCs}+a_{\rm CsI})/2=0.007785{\rm ~nm},
 \end{equation}
 where $a_{\rm CsCs}=0.4685 (Z_{\rm Cs}^{1/2} + Z_{\rm Cs}^{1/2})^{-2/3}=0.007761$ nm and $a_{\rm CsI}=0.4685 (Z_{\rm Cs}^{1/2} + Z_{\rm I}^{1/2})^{-2/3}=0.007809$ nm  correspond to a Cs scattering from a Cs and an I lattice atom, respectively. On the other hand, for Cs recoils from a pure row or plane we use $a_{\rm CsCs}$ because the row or plane from which the recoiling Cs ion was emitted contains  only Cs atoms. Similarly, for I recoils from  a mixed row or plane, we use
\begin{equation}
 \bar{a}_{\rm I}=(a_{\rm II}+a_{\rm CsI})/2=0.007833 {\rm ~nm},
 \end{equation}
 where $a_{\rm II}=0.4685 (Z_{\rm I}^{1/2} + Z_{\rm I}^{1/2})^{-2/3}=0.007857$ nm  and $a_{\rm CsI}$ correspond to an I ion scattering from an I and a Cs lattice atom, respectively. For I recoils from a pure row or plane we use $a_{\rm II}$, since the row or plane the recoiling ion is emitted from is made of I ions only.

To compute the interatomic spacing $d$ in axial directions and the interplanar spacing $d_p$ in planar directions, we have to multiply the lattice constant by the following coefficients~\cite{Gemmell:1974ub}:
\begin{itemize}
  \item Axis: $<100>: 1/2$ , $<110>: 1/\sqrt{2}$ , $<111>: \sqrt{3}/2$
  \item Plane: $\{100\}: 1/2$ , $\{110\}: 1/2\sqrt{2}$ , $\{111\}: 1/2\sqrt{3}$
\end{itemize}

 The Debye temperature  of CsI is $\Theta=125$ K, and the crystals in the KIMS experiment are currently at a temperature of 293 K~\cite{Kims-private}.

\section{Temperature dependence of lattice constant}

In general the lattice constant, $a_{\rm lat}$ is temperature dependent. The change in $a_{\rm lat}$ with temperature depends on the thermal expansion coefficient, $\beta$ of a crystal. For CsI (Tl), $\beta=54 \times 10^{-6}~^\circ$C$^{-1}$. To find the change in the lattice constant at a temperature $T$ and the lattice constant at 20 $^\circ$C, we have
\begin{equation}
[a_{\rm lat}(T) - a_{\rm lat}(20~ ^\circ {\rm C})]/ a_{\rm lat}(20~ ^\circ {\rm C}) = \beta (T - 20~ ^\circ {\rm C}),
\label{alat}
\end{equation}
where $a_{\rm lat}(20~ ^\circ {\rm C})=0.45667$ nm is the lattice constant at 20 $^\circ$C for CsI (Tl). When $T$ changes from 20 $^\circ$C to 600 $^\circ$C, the change in the lattice constant (using Eq.~\ref{alat}) is only 3.1\%. This change in $a_{\rm lat}$ between 20 $^\circ$C and 600 $^\circ$C results in a negligible change in the channeling fractions. As an example the Cs channeling fractions with $c_1=c_2=1$ and $c_1=c_2=2$ for the two choices of  $a_{\rm lat}(20~ ^\circ {\rm C})$ and  $a_{\rm lat}(600~ ^\circ {\rm C})$ are shown in Fig.~\ref{FracCsINaI-alat}(a). As the two curves are very similar, we use $a_{\rm lat}(20~ ^\circ {\rm C})$ for all three crystal temperatures in this paper.
\FIGURE{\epsfig{file=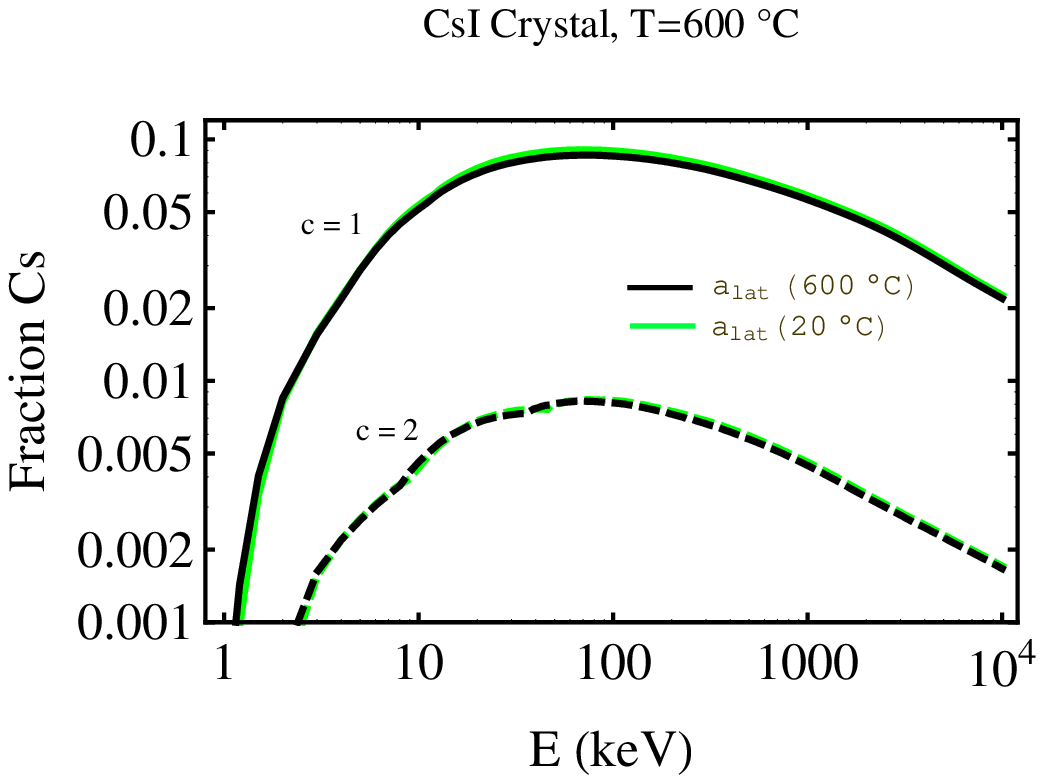,height=160pt}
\epsfig{file=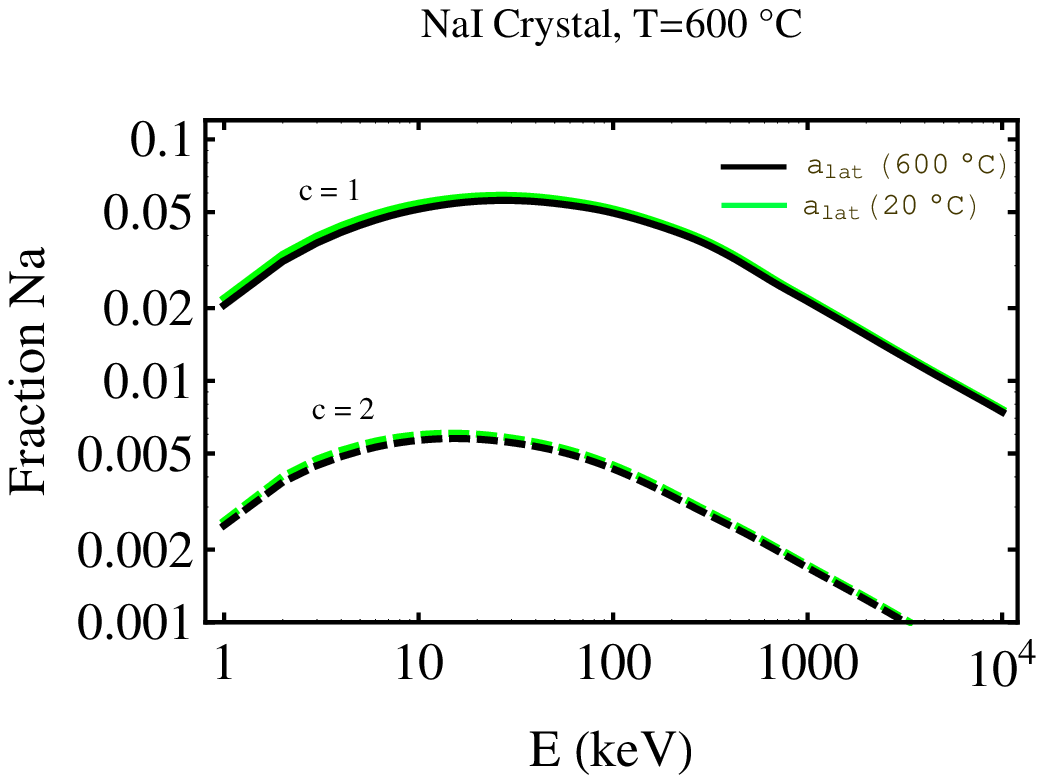,height=160pt}\\
        \vspace{-0.5cm}\caption{Channeling fraction of (a) Cs ions propagating in a CsI crystal and (b) Na ions propagating in an NaI crystal as a function of the recoil energy $E$ for $T=600$ $^\circ$C with $a_{\rm lat}(600~ ^\circ {\rm C})$ (black) and $a_{\rm lat}(20~ ^\circ {\rm C})$ (green/gray) for the two choices of $c_1=c_2=1$ or 2.}%
	\label{FracCsINaI-alat}}
\FIGURE[h]{\epsfig{file=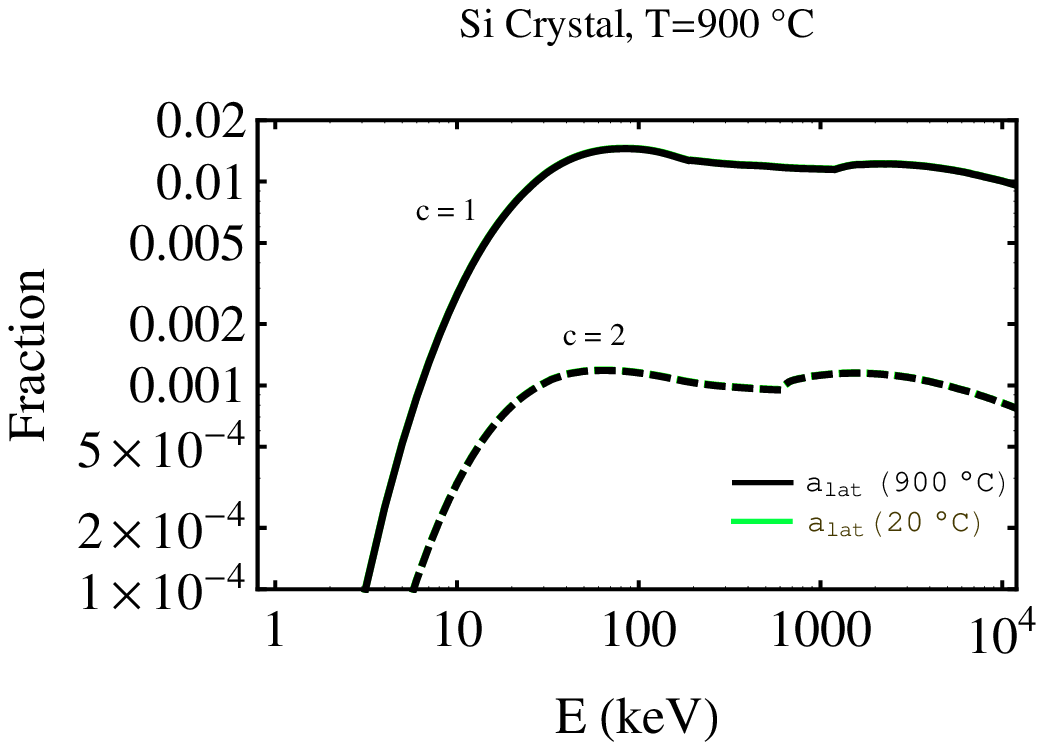,height=155pt}
\epsfig{file=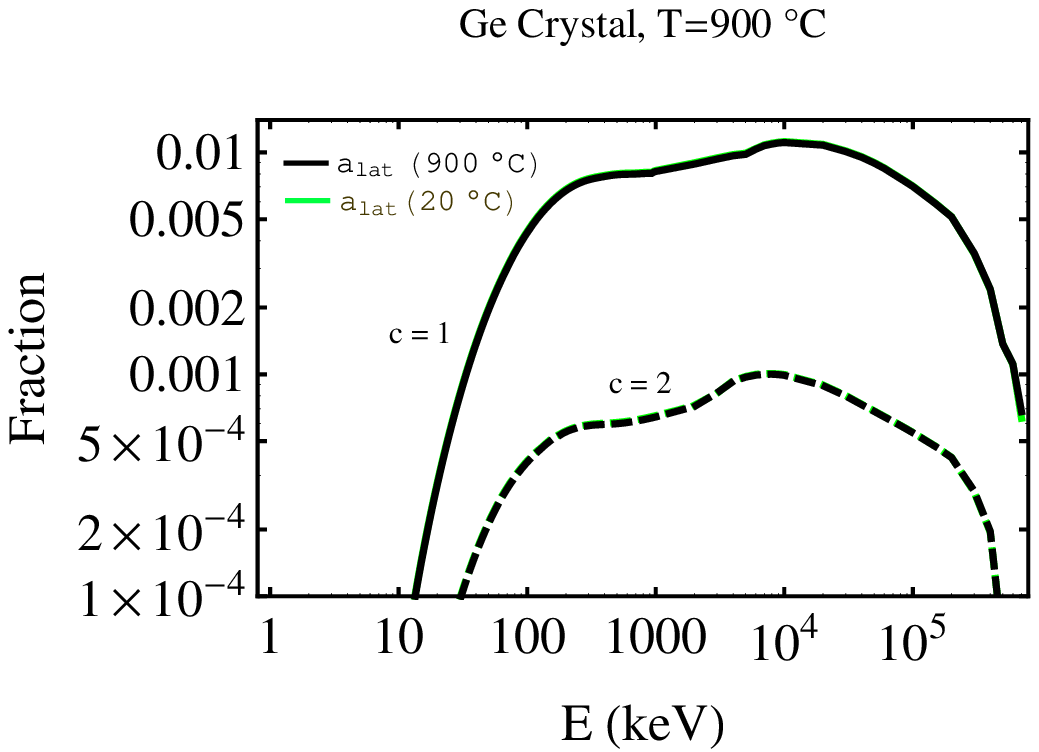,height=155pt}\\
        \vspace{-0.5cm}\caption{Channeling fraction of (a) Si ions propagating in a Si crystal and (b) Ge ions propagating in a Ge crystal as a function of the recoil energy $E$ for $T=900$ $^\circ$C with $a_{\rm lat}(900~ ^\circ {\rm C})$ (black) and $a_{\rm lat}(20~ ^\circ {\rm C})$ (green/gray) for the two choices of $c_1=c_2=1$ or 2.}%
	\label{FracSiGe-alat}}

We can use Eq.~\ref{alat} to find the temperature dependent lattice constant for NaI (Tl), Si and Ge crystals. For these three crystals, the coefficient of thermal expansion is $\beta_{{\rm NaI (Tl)}}=47.4 \times 10^{-6}~^\circ$C$^{-1}$, $\beta_{{\rm Si}}=2.6 \times 10^{-6}~^\circ$C$^{-1}$, and $\beta_{{\rm Ge}}=5.9 \times 10^{-6}~^\circ$C$^{-1}$. When $T$ changes from 20 $^\circ$C to 600 $^\circ$C, the change in the lattice constant of NaI (Tl) is 2.75\%. In Si and Ge, we can go to higher temperatures, and the maximum temperature that we considered in our previous paper on Si and Ge~\cite{BGG-II} was 900 $^\circ$C. When $T$ changes from 20 $^\circ$C to 900 $^\circ$C, the change in the lattice constant of Si and Ge is 0.23\% and 0.52\% respectively.

The Na channeling fractions with $c_1=c_2=1$ and $c_1=c_2=2$ for the two choices of  $a_{\rm lat}(20~ ^\circ {\rm C})$ and  $a_{\rm lat}(600~ ^\circ {\rm C})$ are shown in Fig.~\ref{FracCsINaI-alat}(b) for an NaI crystal. Fig.~\ref{FracSiGe-alat} shows the channeling fractions for Si and Ge for the two choices of $a_{\rm lat}(20~ ^\circ {\rm C})$ and  $a_{\rm lat}(900~ ^\circ {\rm C})$.

Clearly, the change in the curves is negligible. Thus we always used the value of $a_{\rm lat}$ measured at 20 $^\circ$C in this paper as well as in our previous papers~\cite{BGG-I, BGG-II}.

\end{document}